# A Bayesian Probability Calculus for Density Matrices*

**Manfred K. Warmuth and Dima Kuzmin**
Computer Science Department
University of California, Santa Cruz
manfred|dima@cse.ucsc.edu

## Abstract

One of the main concepts in quantum physics is a density matrix, which is a symmetric positive definite matrix of trace one. Finite probability distributions can be seen as a special case when the density matrix is restricted to be diagonal.

We develop a probability calculus based on these more general distributions that includes definitions of joints, conditionals and formulas that relate these, including analogs of the Theorem of Total Probability and various Bayes rules for the calculation of posterior density matrices. The resulting calculus parallels the familiar "conventional" probability calculus and always retains the latter as a special case when all matrices are diagonal.

Whereas the conventional Bayesian methods maintain uncertainty about which model has the highest data likelihood, the generalization maintains uncertainty about which unit direction has the largest variance. Surprisingly the bounds also generalize: as in the conventional setting we upper bound the negative log likelihood of the data by the negative log likelihood of the MAP estimator.

## 1 Introduction

The main notion of a "mixture state" used in quantum physics is a density matrix. States are unit vectors. For the sake of simplicity we assume in this paper that the underlying vector space is $\mathbb{R}^n$ (for finite $n$). Each state $\boldsymbol{u}$ (unit column vector in $\mathbb{R}^n$) is associated with a *dyad* $\boldsymbol{u}\boldsymbol{u}^T \in \mathbb{R}^{n\times n}$, which is a degenerate ellipse with a single axis in direction $\pm\boldsymbol{u}$ that has radius one (See Figure 1). The dyad $\boldsymbol{u}\boldsymbol{u}^\top$ may also be seen as a one-dimensional projection matrix which projects any vector onto direction $\boldsymbol{u}$. These dyads are the elementary events of a *generalized probability space*. It is useful to keep the corresponding "conventional" space in mind, which consists of a finite set of size $n$. The $n$ points are the elementary events and a probability distribution may be seen as a mixture over the $n$ points. In the generalized case there are infinitely many dyads even if the dimension $n$ is finite.[1]

Density matrices generalize finite probability distributions. They can be defined as mixtures of dyads $\boldsymbol{A} = \sum_i \alpha_i \boldsymbol{a}_i\boldsymbol{a}_i^T$ where the mixture coefficients $\alpha_i$ are non-negative and sum to one. Such mixtures are symmetric[2] and positive definite. Dyads have trace one:

$$\operatorname{tr}(\boldsymbol{u}\boldsymbol{u}^\top) = \operatorname{tr}(\boldsymbol{u}^\top\boldsymbol{u}) = \|\boldsymbol{u}\|_2^2 = 1.$$

Therefore, density matrices also have trace one. There may be an arbitrary number of unit vectors $\boldsymbol{a}_i$ in the above sum. However, any density matrix can be decomposed into mixture of $n$ orthogonal dyads, one for each eigenvector (See Figure 1).

A density matrix $\boldsymbol{A}$ assigns generalized probability $\operatorname{tr}(\boldsymbol{A}\boldsymbol{u}\boldsymbol{u}^\top)$ to each unit vector $\boldsymbol{u}$ and its associated dyad $\boldsymbol{u}\boldsymbol{u}^T$ (see Figure 2). This probability is independent of how $\boldsymbol{A}$ is expressed as a mixture and can be rewritten as $\operatorname{tr}(\boldsymbol{u}^T\boldsymbol{A}\boldsymbol{u}) = \boldsymbol{u}^T\boldsymbol{A}\boldsymbol{u}$. Note that if $\boldsymbol{A}$ is viewed as a covariance matrix of a random cost vector $\boldsymbol{c}$, then $\boldsymbol{u}^T\boldsymbol{A}\boldsymbol{u}$ is the variance of the cost along direction $\boldsymbol{u}$, i.e. the variance of $\boldsymbol{c}\cdot\boldsymbol{u}$.

If we view a conventional mixture as a diagonal matrix where the mixture coefficients $\alpha_i$ form the diagonal, then the diagonal matrix can be written as $\sum_i \alpha_i \boldsymbol{e}_i\boldsymbol{e}_i^T$, where $\boldsymbol{e}_i$ are the standard basis vectors. This means that conventional distributions are special density matrices where the eigensystem is restricted to be the identity matrix. In this paper develop a Bayesian style analysis for the case when the eigensystem is allowed to be arbitrary.

* Supported by NSF grant CCR 9821087. Some of this work was done while visiting National ICT Australia in Canberra.

[1] The machinery for infinite dimensional vector spaces is available. However, in this paper we start with the simplest finite dimensional setting.

[2] In quantum physics complex numbers are used instead of reals. In that case "symmetric" is replaced by "hermitian" and all our formulas hold for that case as well.

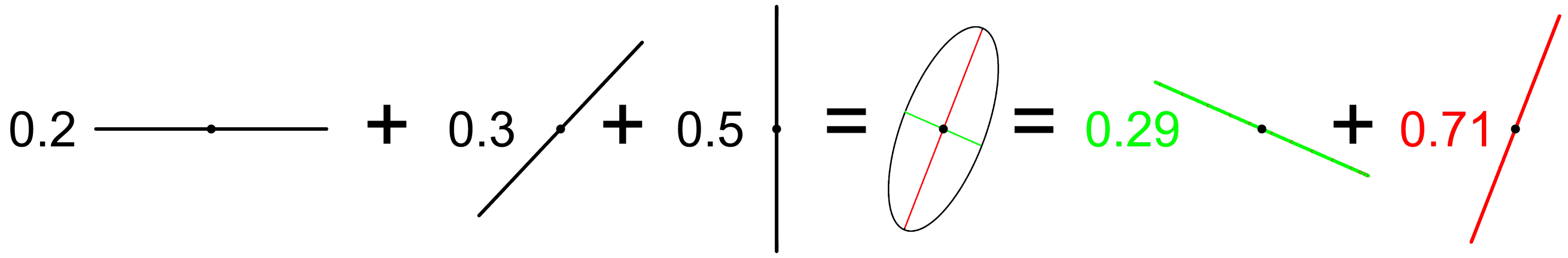

Figure 1: Two different dyad mixtures that lead to the same density matrix:
$0.2 \left(\begin{smallmatrix}1\\0\end{smallmatrix}\right)(1\ 0) + 0.3 \begin{pmatrix}\sqrt{2}/2\\ \sqrt{2}/2\end{pmatrix}\left(\frac{\sqrt{2}}{2}\ \frac{\sqrt{2}}{2}\right) + 0.5\left(\begin{smallmatrix}0\\1\end{smallmatrix}\right)(0\ 1) = \left(\begin{smallmatrix}0.35 & 0.15\\ 0.15 & 0.65\end{smallmatrix}\right) = 0.29\left(\begin{smallmatrix}-0.92\\ 0.38\end{smallmatrix}\right)(-0.92\ 0.38) + 0.71\left(\begin{smallmatrix}0.38\\ 0.92\end{smallmatrix}\right)(0.38\ 0.92)$.
Matrices are depicted as ellipses and dyads are degenerate single axis ellipses.

Perhaps the simplest case to see that something unusual is going on is the uniform density matrix, i.e. $\frac{1}{n}$ times identity $\boldsymbol{I}$. This density matrix assigns probability $\frac{1}{n}$ to every unit vector, even though there are infinitely many of them. However note that the sum of generalized probabilities over any $n$ orthogonal dyads $\boldsymbol{a}_i\boldsymbol{a}_i^T$ is one (and this holds if $\frac{1}{n}\boldsymbol{I}$ is replaced by any density matrix):

$$\sum_{i=1}^{n} \operatorname{tr}\left(\frac{1}{n}\boldsymbol{I}\ \boldsymbol{a}_i\boldsymbol{a}_i^T\right) = \operatorname{tr}\left(\frac{1}{n}\boldsymbol{I}\ \underbrace{\sum_i \boldsymbol{a}_i\boldsymbol{a}_i^T}_{\boldsymbol{I}}\right) = \operatorname{tr}\left(\frac{1}{n}\boldsymbol{I}\right) = 1.$$

Note that while in the conventional case probabilities are additive over the points in the set, in the generalized case probabilities are additive over orthogonal sets of dyads.

In this paper we use density matrices as generalized priors and develop a unifying Bayesian probability calculus for density matrices with rules for translating between joints and conditionals. All formulas retain the conventional case as the special case when the matrices are diagonal. In previous work [War05] we derived a generalized Bayes rule based on the minimum relative entropy principle, but no satisfactory probabilistic interpretation was given for this rule. This Bayes rule fits nicely into our new calculus and we can interpret it using the notion of generalized probability introduced above. Also, in the context of quantum information theory, a formula was proposed for the conditional density between two probability spaces [CA99]. Again this special formula is put in a more general context.

For any fixed orthonormal eigensystem $\boldsymbol{a}_i$ one can use the dyads $\boldsymbol{a}_i\boldsymbol{a}_i^T$ as elementary events of a conventional probability space. As already discussed any density matrix assigns conventional probabilities to these events (that sum to one). Our approach is fundamentally different in that we use density matrices to maintain uncertainty over all eigensystems. Our conditional density matrices are part of the probabilistic system specified by a generalized joint probability distribution. In particular, our conditioning method leads to generalizations of the Theorem of Total Probability that involve density matrices.

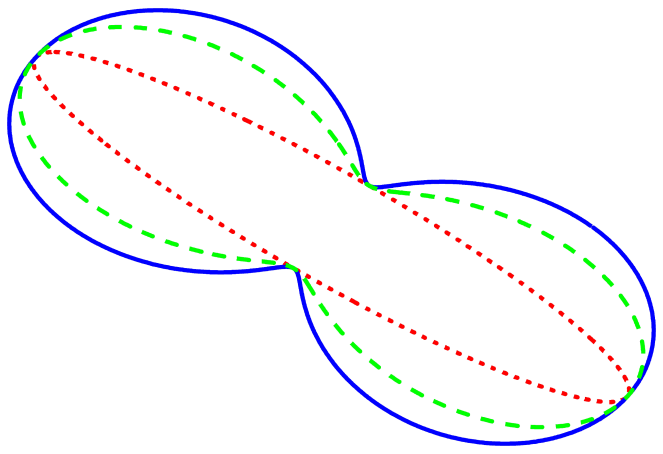

Figure 2: The ellipse is a plot of $\{\boldsymbol{Au} : \boldsymbol{u} \text{ unit}\}$, for some density matrix $\boldsymbol{A}$. The outermost solid figure-eight is a plot of the generalized probabilities in direction $\boldsymbol{u}$, i.e. $\operatorname{tr}(\boldsymbol{Auu}^\top)\boldsymbol{u}$. The inner dashed figure-eight depicts the quantity $\operatorname{tr}(\boldsymbol{A} \odot \boldsymbol{uu}^\top)\boldsymbol{u}$. Note that the dashed figure-eight is contained in the solid figure-eight and this illustrates the inequality given in property 3 of the $\odot$ operation.

Our approach also differs from the work of [SBC01] who derive a "quantum Bayes rule" that describes uncertainty information about unobserved quantum measurements of a composite system as a density matrix. At this point we don't have a quantum physical interpretation of our calculus and the Bayes rules it contains. Our work is based on a variational derivation of a generalized Bayes rule that parallels the derivation of the conventional Bayes rule [War05].

## 2 Generalized Probability Distributions and Density Matrices

For the sake of simplicity we assume that our vector space is $\mathbb{R}^n$. However, everything discussed in this section holds for separable finite or infinite dimensional real and complex Hilbert spaces.

A function $\mu(\boldsymbol{u})$ from unit vectors $\boldsymbol{u}$ in $\mathbb{R}^n$ to $\mathbb{R}$ is called a *generalized probability measure* if the following

two conditions hold:

- $\forall \boldsymbol{u}$, $0 \leq \mu(\boldsymbol{u}) \leq 1$.
- If $\boldsymbol{u}_1, \ldots, \boldsymbol{u}_n$ form an orthonormal basis for $\mathbb{R}^n$, then $\sum \mu(\boldsymbol{u}_i) = 1$.

Gleason's Theorem states that there is a one-to-one correspondence between generalized probability measures and density matrices[3] in $\mathbb{R}^{n\times n}$:

**Theorem 2.1** *[Gle57] Let $n \geq 3$.[4] Then any generalized probability measure $\mu$ on $\mathbb{R}^n$ has the form $\mu(\boldsymbol{u}) = \mathrm{tr}(\boldsymbol{A}\boldsymbol{u}\boldsymbol{u}^\top)$, for a uniquely defined density matrix $\boldsymbol{A}$.*

It is easy to see that every density matrix defines a generalized probability measure. The other direction, however, is highly non-trivial. As discussed in the introduction, the dyads $\boldsymbol{u}\boldsymbol{u}^T$ function as elementary events. One may ask what corresponds to arbitrary events and how probabilities are defined for them. In the conventional case, an event is a subset of the domain which can be represented as a vector in $\{0,1\}^n$. In the generalized setting, an event is a symmetric positive definite matrix $\boldsymbol{S}$ with eigenvalues $\sigma_i \in \{0,1\}$. Each such matrix $\boldsymbol{S}$ with eigendecomposition $\sum_i \sigma_i \boldsymbol{s}_i \boldsymbol{s}_i^T$ is a projection matrix for a subspace of $\mathbb{R}^n$ and its probability w.r.t. a density $\boldsymbol{A}$ is defined as a weighted sum of the probabilities of the elementary events defining $\boldsymbol{S}$ (Recall that the $\mathrm{tr}(\boldsymbol{A}\,\boldsymbol{s}_i\boldsymbol{s}_i^T)$ sum to one):

$$\mathrm{tr}(\boldsymbol{A}\boldsymbol{S}) = \sum_i \sigma_i \mathrm{tr}(\boldsymbol{A}\,\boldsymbol{s}_i\boldsymbol{s}_i^T).$$

Random variables are defined in an analogous way. In the conventional case a random variable associates a real value with each point. Now a random variable is an arbitrary symmetric matrix $\boldsymbol{S}$. Such matrices have arbitrary real numbers as their eigenvalues and the trace $\mathrm{tr}(\boldsymbol{A}\boldsymbol{S})$ as expanded above becomes the expectation of the random variable w.r.t. density $\boldsymbol{A}$. When $\boldsymbol{S}$ is an event, then this expectation is also a probability and lies in [0..1]. As discussed in the introduction, the conventional case is always retained as a special case when all the matrices are diagonal (i.e. fixed eigensystem $\boldsymbol{I}$).

One might ask if the generalized probability measures can be somehow reduced to conventional probabilities. A natural interpretation of a density matrix is to view it as a parameterized distribution over the unit sphere: if $\mu(\boldsymbol{u})$ is a uniform measure on the sphere then for any positive definite matrix $\boldsymbol{A} \in \mathbb{R}^{n\times n}$ of trace $n$, $\boldsymbol{u}^T\boldsymbol{A}\,\boldsymbol{u}\,\mu(\boldsymbol{u})$ is a conventional probability density on the sphere:

$$\int \boldsymbol{u}^T \overbrace{\sum_i \alpha_i \boldsymbol{a}_i \boldsymbol{a}_i^T}^{\boldsymbol{A}} \boldsymbol{u}\,\mu(\boldsymbol{u})d\boldsymbol{u} = \sum_i \alpha_i \int (\boldsymbol{u}^T\boldsymbol{a}_i)^2 \mu(\boldsymbol{u})d\boldsymbol{u}$$

$$= \mathrm{tr}(\boldsymbol{A}) \int (\boldsymbol{u}^T(\frac{1}{\sqrt{n}}))^2 \mu(\boldsymbol{u})d\boldsymbol{u} = \frac{\mathrm{tr}(\boldsymbol{A})}{n} \underbrace{\int \overbrace{(\boldsymbol{u})^2}^{1} \mu(\boldsymbol{u})d\boldsymbol{u}}_{1}.$$

This interprets density matrices in a conventional way, but one still needs to assign probabilities to the elementary events. In particular, the value of density function on a point is not a probability and any unit vector gets probability zero under the measure just defined. Essentially, conventional probability spaces cannot satisfactorily model generalized probabilities, but the details are rather involved. See [Hol01] for an extended discussion.

## 3 Joint Distributions

So far we talked about a single generalized probability space. Now we need to consider several spaces and joint distributions over them. In the conventional case $A, B$ denote finite sets, $P(A), P(B)$ probability vectors over these sets and $P(A,B)$ is a $n_A \times n_B$ dimensional matrix of probabilities for the tuple set $A \times B$. In the generalized case, $\mathbb{A}, \mathbb{B}$ denote real finite dimensional vector spaces and $\boldsymbol{D}(\mathbb{A}), \boldsymbol{D}(\mathbb{B})$ are the density matrices defining the generalized probability distributions over these spaces. The *joint space* $(\mathbb{A}, \mathbb{B})$ is the tensor product[5] between the spaces $\mathbb{A}$ and $\mathbb{B}$, which is of dimension $n_{\mathbb{A}} n_{\mathbb{B}}$. The density matrix associated with the joint space is denoted by $\boldsymbol{D}(\mathbb{A}, \mathbb{B})$.

We let $\boldsymbol{D}(\boldsymbol{a}), \boldsymbol{D}(\boldsymbol{b})$ denote the probabilities assigned to unit vectors $\boldsymbol{a}$ and $\boldsymbol{b}$ from the spaces $\mathbb{A}$ and $\mathbb{B}$ by the density matrices $\boldsymbol{D}(\mathbb{A})$ and $\boldsymbol{D}(\mathbb{B})$, respectively, i.e. $\boldsymbol{D}(\boldsymbol{a}) = \mathrm{tr}(\boldsymbol{D}(\mathbb{A})\boldsymbol{a}\boldsymbol{a}^\top)$ and $\boldsymbol{D}(\boldsymbol{b}) = \mathrm{tr}(\boldsymbol{D}(\mathbb{B})\boldsymbol{b}\boldsymbol{b}^\top)$. The conventional probability distributions are diagonal density matrices. A probability distribution $P(A)$ on the set $A$ is the density matrix $\mathrm{diag}(P(A))$. Also $P(a_i) = \boldsymbol{e}_i^T \mathrm{diag}(P(A))\boldsymbol{e}_i$.

To introduce the joint probability $\boldsymbol{D}(\boldsymbol{a}, \boldsymbol{b})$ we need the Kronecker matrix product. Given two matrices $\boldsymbol{E}$ and $\boldsymbol{F}$ of dimensions $n_{\boldsymbol{E}} \times n_{\boldsymbol{E}}$ and $n_{\boldsymbol{F}} \times n_{\boldsymbol{F}}$, their Kronecker product (also known as the direct product or tensor product) $\boldsymbol{E} \otimes \boldsymbol{F}$ is a matrix of dimension $n_{\boldsymbol{E}} n_{\boldsymbol{F}} \times n_{\boldsymbol{E}} n_{\boldsymbol{F}}$, which in block form is a $n_{\boldsymbol{E}} \times n_{\boldsymbol{E}}$ block matrix with the $ij$th block given by $e_{ij}\boldsymbol{F}$:

$$\boldsymbol{E} \otimes \boldsymbol{F} = \begin{pmatrix} e_{11}\boldsymbol{F} & \dots & e_{1n_{\boldsymbol{E}}}\boldsymbol{F} \\ \dots & \dots & \dots \\ e_{n_{\boldsymbol{E}}1}\boldsymbol{F} & \dots & e_{n_{\boldsymbol{E}}n_{\boldsymbol{E}}}\boldsymbol{F} \end{pmatrix}$$

[3] The core of the original proof of Gleason's Theorem was for $\mathbb{R}^3$ [Gle57], and he then generalized the proof to separable real and complex Hilbert spaces of dimension $n \geq 3$.

[4] A slightly different version of this theorem that is based on *effects* instead of dyads holds for dimension 2 as well [CFMR04].

[5] See [Bha97] for a formal definition of tensor product between vector spaces. For us, the tensor product of $\mathbb{R}^{n_{\mathbb{A}}}$ and $\mathbb{R}^{n_{\mathbb{B}}}$ is $\mathbb{R}^{n_{\mathbb{A}} n_{\mathbb{B}}}$.

A few useful properties of Kronecker products:

1. $(\boldsymbol{E} \otimes \boldsymbol{F})(\boldsymbol{G} \otimes \boldsymbol{H}) = \boldsymbol{E}\boldsymbol{G} \otimes \boldsymbol{F}\boldsymbol{H}$ if the dimensions are compatible.
2. $\mathrm{tr}(\boldsymbol{E} \otimes \boldsymbol{F}) = \mathrm{tr}(\boldsymbol{E})\mathrm{tr}(\boldsymbol{F})$.
3. If $\boldsymbol{F}$ has eigenvalues $\phi_i$ and eigenvectors $\boldsymbol{f}_i$ and similarly $\boldsymbol{G}$ has eigenvalues $\gamma_j$ and eigenvectors $\boldsymbol{g}_j$, then $\boldsymbol{F} \otimes \boldsymbol{G}$ has eigenvalues $\phi_i\gamma_j$ and eigenvectors $\boldsymbol{f}_i \otimes \boldsymbol{g}_j$.

Now the joint probability $\boldsymbol{D}(\boldsymbol{a}, \boldsymbol{b})$ becomes the probability assigned by density matrix $\boldsymbol{D}(\mathbb{A}, \mathbb{B})$ to the jointly specified unit vector $\boldsymbol{a} \otimes \boldsymbol{b}$:

$$\begin{aligned} \boldsymbol{D}(\boldsymbol{a}, \boldsymbol{b}) &= \mathrm{tr}(\boldsymbol{D}(\mathbb{A}, \mathbb{B})(\boldsymbol{a} \otimes \boldsymbol{b})(\boldsymbol{a} \otimes \boldsymbol{b})^\top) \\ &= \mathrm{tr}(\boldsymbol{D}(\mathbb{A}, \mathbb{B})(\boldsymbol{a}\boldsymbol{a}^\top \otimes \boldsymbol{b}\boldsymbol{b}^\top)). \end{aligned}$$

Note that in the conventional case a joint between two sets $A$ and $B$ is defined over all pairs of points from $A$ and $B$. However, in the generalized case, there are elementary events in the joint that don't decompose into elementary events for the marginals: there are dyads in the joint space that are not of the form $(\boldsymbol{a} \otimes \boldsymbol{b})(\boldsymbol{a} \otimes \boldsymbol{b})^\top$. This possibility leads to non-conventional correlations in our joints, or what quantum physicists call "entanglement".

# 4 Marginalization via Partial Traces

We would like to be able to perform marginalization operations on our joint density matrix $\boldsymbol{D}(\mathbb{A}, \mathbb{B})$. In the conventional case, a marginalization was performed by summing out one of the variables. For density matrices, the analog is the *partial trace* (see e.g. [NC00]). Suppose we have some matrix $\boldsymbol{G}$ defined on the joint space $(\mathbb{A}, \mathbb{B})$ with dimensionality $n_\mathbb{A} n_\mathbb{B} \times n_\mathbb{A} n_\mathbb{B}$. Let's write $\boldsymbol{G}$ in block form as a $n_\mathbb{A} \times n_\mathbb{A}$ matrix of $n_\mathbb{B} \times n_\mathbb{B}$ blocks $\boldsymbol{G}_{ij}$. Then the two partial traces of $\boldsymbol{G}$ are given as:

$$\underbrace{\mathrm{tr}_\mathbb{A}(\boldsymbol{G})}_{n_\mathbb{B} \times n_\mathbb{B}} = \boldsymbol{G}_{11} + \boldsymbol{G}_{22} + \ldots + \boldsymbol{G}_{n_\mathbb{A} n_\mathbb{A}}$$

$$\underbrace{\mathrm{tr}_\mathbb{B}(\boldsymbol{G})}_{n_\mathbb{A} \times n_\mathbb{A}} = \begin{pmatrix} \mathrm{tr}(\boldsymbol{G}_{11}) & \dots & \mathrm{tr}(\boldsymbol{G}_{1n_\mathbb{A}}) \\ \dots & \dots & \dots \\ \mathrm{tr}(\boldsymbol{G}_{n_\mathbb{A}1}) & \dots & \mathrm{tr}(\boldsymbol{C}_{n_\mathbb{A} n_\mathbb{A}}) \end{pmatrix}$$

Properties of partial traces ($\boldsymbol{E}$ is a matrix on $\mathbb{A}$, $\boldsymbol{F}$ is a matrix on $\mathbb{B}$):

1. $\mathrm{tr}_\mathbb{A}(\boldsymbol{E} \otimes \boldsymbol{F}) = \mathrm{tr}(\boldsymbol{E})\boldsymbol{F}, \quad \mathrm{tr}_\mathbb{B}(\boldsymbol{E} \otimes \boldsymbol{F}) = \mathrm{tr}(\boldsymbol{F})\boldsymbol{E}$
2. $\mathrm{tr}(\boldsymbol{G}) = \mathrm{tr}(\mathrm{tr}_\mathbb{A}(\boldsymbol{G})) = \mathrm{tr}(\mathrm{tr}_\mathbb{B}(\boldsymbol{G}))$
3. $\mathrm{tr}_\mathbb{A}(\boldsymbol{G}(\boldsymbol{I}_\mathbb{A} \otimes \boldsymbol{F})) = \mathrm{tr}_\mathbb{A}(\boldsymbol{G})\boldsymbol{F}$
   $\mathrm{tr}_\mathbb{A}((\boldsymbol{I}_\mathbb{A} \otimes \boldsymbol{F})\boldsymbol{G}) = \boldsymbol{F}\mathrm{tr}_\mathbb{A}(\boldsymbol{G})$
4. $\mathrm{tr}(\boldsymbol{G}(\boldsymbol{E} \otimes \boldsymbol{F})) = \mathrm{tr}(\mathrm{tr}_\mathbb{B}(\boldsymbol{G}(\boldsymbol{I}_\mathbb{A} \otimes \boldsymbol{F}))\boldsymbol{E})$ .

It can be shown that the partial trace of a density matrix is also a density matrix (on the reduced space). Thus we can use $\boldsymbol{D}(\mathbb{A}) = \mathrm{tr}_\mathbb{B}(\boldsymbol{D}(\mathbb{A}, \mathbb{B}))$ as the marginalization. Via partial traces we can also get objects of the type $\boldsymbol{D}(\mathbb{A}, \boldsymbol{b})$, which is a symmetric positive definite matrix of trace at most one. In the conventional case it corresponds to taking one row out of the probability table. In the generalized case we want the following property to be satisfied: $\mathrm{tr}(\boldsymbol{D}(\mathbb{A}, \boldsymbol{b})\boldsymbol{a}\boldsymbol{a}^\top) = \boldsymbol{D}(\boldsymbol{a}, \boldsymbol{b})$. This is accomplished by defining $\boldsymbol{D}(\mathbb{A}, \boldsymbol{b})$ as $\mathrm{tr}_\mathbb{B}(\boldsymbol{D}(\mathbb{A}, \mathbb{B})(\boldsymbol{I}_A \otimes \boldsymbol{b}\boldsymbol{b}^\top))$. The desired condition now follows from partial trace property 4. The above properties also imply that $\mathrm{tr}(\boldsymbol{D}(\mathbb{A}, \boldsymbol{b})) = \boldsymbol{D}(\boldsymbol{b})$.

Now we can also give the definition of independence: $\boldsymbol{D}(\mathbb{A})$ is independent of $\boldsymbol{D}(\mathbb{B})$ if the joint density matrix decomposes: $\boldsymbol{D}(\mathbb{A}, \mathbb{B}) = \boldsymbol{D}(\mathbb{A}) \otimes \boldsymbol{D}(\mathbb{B})$.

# 5 Commutative Matrix Product Operation

It is well-known that the product of two symmetric positive definite matrices can be neither symmetric nor positive definite (See Figure 3). As in [War05], we introduce the $\odot$ operation which generalizes the matrix product and does not have these drawbacks. It will later be used to define conditional density matrices and generalizations of the Bayes rule. For symmetric and strictly positive definite matrices we define $\odot$ as:

$$\boldsymbol{S} \odot \boldsymbol{T} := \mathbf{exp}(\mathbf{log}\, \boldsymbol{S} + \mathbf{log}\, \boldsymbol{T}), \tag{1}$$

where here the exponential and logarithms are matrix functions. For the exponential of any symmetric matrix $\boldsymbol{R} = \sum_i \rho_i \boldsymbol{r}_i \boldsymbol{r}_i^T$ we simply exponentiate the eigenvalues of the matrix, i.e. $\mathbf{exp}(\boldsymbol{R}) := \sum_i \exp(\rho_i)\boldsymbol{r}_i\boldsymbol{r}_i^T$ The logarithm of a matrix is defined analogously. For arbitrary matrices, these operations can be defined by a series.

The matrix log of both matrices produces symmetric matrices which are closed under addition and the matrix exponential of the sum returns a symmetric positive definite matrix. See Figure 3 for a comparison of matrix product and $\odot$. Although the matrix log is not defined when the matrix has a zero eigenvalue, one can define the operation $\odot$ for arbitrary symmetric positive definite matrices as the following limit:

$$\boldsymbol{S} \odot \boldsymbol{T} := \lim_{n \to \infty} (\boldsymbol{S}^{1/n}\boldsymbol{T}^{1/n})^n.$$

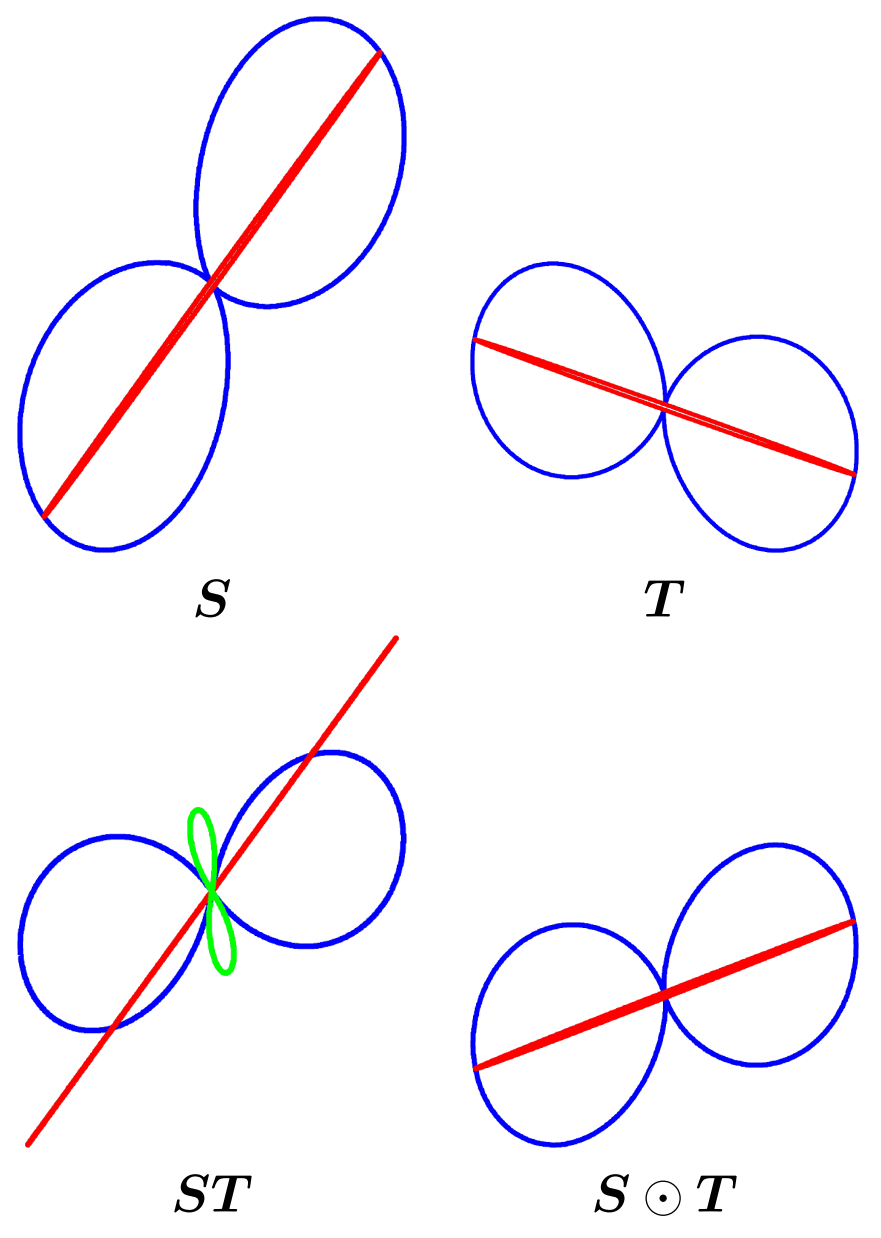

Figure 3: Comparison between regular matrix product and the $\odot$ operation. As before we plot the ellipse $\boldsymbol{S}\boldsymbol{u}$ and the figure eight $\mathrm{tr}(\boldsymbol{S}\boldsymbol{u}\boldsymbol{u}^\top)\,\boldsymbol{u}$ for each matrix. Note that the ellipses are very thin and look almost like lines. The "ears" in the plot for $\boldsymbol{ST}$ mean that $\mathrm{tr}(\boldsymbol{ST}\boldsymbol{u}\boldsymbol{u}^\top)$ can get negative, and thus $\boldsymbol{ST}$ is not positive definite.

This limit is the *Lie Product Formula*[6] [Bha97] when $\boldsymbol{S}$ and $\boldsymbol{T}$ are both strictly positive definite and symmetric, but it exists even if the matrices don't have full rank. As $n$ increases, $(\boldsymbol{S}^{1/n}\boldsymbol{T}^{1/n})^n$ gets closer and closer to being positive definite and symmetric. The first couple iteration of the limit formula are plotted in Figure 4. See [Ale02] for additional plots.

By Theorem 1.2 of [Sim79] we have that:

$$\mathrm{range}(\boldsymbol{S}\odot\boldsymbol{T}) \;=\; \mathrm{range}(\boldsymbol{S})\cap\mathrm{range}(\boldsymbol{T}),$$

where the range of a matrix is the linear subspace spanned by the columns of the matrix. This property generalizes the intersection property underlying the conventional setting, where the element-wise product of the characteristic vectors of two subsets gives a characteristic vector for the subset intersection.

Let $\boldsymbol{B_S}$ be a matrix whose columns form an orthonormal basis for the range of $\boldsymbol{S}$, i.e. $\boldsymbol{B_S}\in\mathbb{R}^{n\times k}$ and $\boldsymbol{B_S}^\top\boldsymbol{B_S} = \boldsymbol{I}_k$, where $k$ is the dimensionality of the range of $\boldsymbol{S}$. Define $\boldsymbol{B_T}$ analogously. In a similar fashion $\boldsymbol{B_{S\cap T}}$ will contain the basis for the intersection of ranges. Let $\mathbf{log}^+$ denote the modified matrix logarithm that takes the log of non-zero eigenvalues but leaves the zero eigenvalues unchanged. We can now re-express the $\odot$ operation as

$$\boldsymbol{S}\odot\boldsymbol{T}=\boldsymbol{B_{S\cap T}}\,\mathbf{exp}(\boldsymbol{B}^\top_{\boldsymbol{S\cap T}}(\mathbf{log}^+\boldsymbol{S}+\mathbf{log}^+\boldsymbol{T})\boldsymbol{B_{S\cap T}})\boldsymbol{B}^\top_{\boldsymbol{S\cap T}}.$$

The $\odot$ operation has many interesting properties, some of which are listed below. Again, $\boldsymbol{S}$ and $\boldsymbol{T}$ are arbitrary symmetric positive definite matrices:

1. $\odot$ is commutative, associative (with identity matrix being the neutral element) and preserves symmetry and positive definiteness.

2. $\boldsymbol{S}\odot\boldsymbol{T} = \boldsymbol{ST}$ iff $\boldsymbol{S}$ and $\boldsymbol{T}$ commute.

3. $\mathrm{tr}(\boldsymbol{S}\odot\boldsymbol{T}) \le \mathrm{tr}(\boldsymbol{ST})$ with equality when $\boldsymbol{S}$ and $\boldsymbol{T}$ commute. See Figure 2 for an illustration of the inequality. For strictly positive definite $\boldsymbol{S}$ and $\boldsymbol{T}$ this inequality is an instantiation of the well-known Golden-Thompson Inequality[7] (See e.g. [Bha97]).

4. For any unit direction $\boldsymbol{u}\in\mathrm{range}(\boldsymbol{S})$, $\boldsymbol{u}\boldsymbol{u}^\top\odot\boldsymbol{S} = e^{\boldsymbol{u}^\top(\mathbf{log}^+\boldsymbol{S})\boldsymbol{u}}\,\boldsymbol{u}\boldsymbol{u}^\top$.

5. $\det(\boldsymbol{S}\odot\boldsymbol{T}) = \det(\boldsymbol{S})\det(\boldsymbol{T})$, the same as for the regular matrix product.

## 6 Conditionals i.t.o. Joints

The topic of conditional probabilities in this generalized setting contains many subtleties. First we will give the defining formulas for conditional density matrices and then briefly discuss some of the issues.

1. $\boldsymbol{D}(\mathbb{A}|\mathbb{B}) := \boldsymbol{D}(\mathbb{A},\mathbb{B})\odot(\boldsymbol{I}_{\mathbb{A}}\otimes\boldsymbol{D}(\mathbb{B}))^{-1}$ (Formula (4) of [CA99] expressed with the $\odot$ operation).

2. $\boldsymbol{D}(\mathbb{A}|\boldsymbol{b}) := \frac{\boldsymbol{D}(\mathbb{A},\boldsymbol{b})}{\mathrm{tr}(\boldsymbol{D}(\mathbb{A},\boldsymbol{b}))}$.

3. $\boldsymbol{D}(\boldsymbol{a}|\mathbb{B}) := \boldsymbol{D}(\boldsymbol{a},\mathbb{B})\odot\boldsymbol{D}(\mathbb{B})^{-1}$.

4. $\boldsymbol{D}(\boldsymbol{a}|\boldsymbol{b}) := \frac{\boldsymbol{D}(\boldsymbol{a},\boldsymbol{b})}{\boldsymbol{D}(\boldsymbol{b})}$. This basic conditional probability is a straightforward generalization of the conventional case. In the full paper we show how to set up a quantum measurement on the joint density matrix $\boldsymbol{D}(\mathbb{A},\mathbb{B})$ s.t. a particular outcome has probability $\boldsymbol{D}(\boldsymbol{a}|\boldsymbol{b})$.

To complete the rules for conditional density matrices, we would need rules that allow us to marginalize the conditionals, e.g. for going from $\boldsymbol{D}(\mathbb{A}|\mathbb{B})$ to $\boldsymbol{D}(\boldsymbol{a}|\boldsymbol{b})$. One obvious rule is $\boldsymbol{D}(\boldsymbol{a}|\boldsymbol{b}) = \mathrm{tr}(\boldsymbol{D}(\mathbb{A}|\boldsymbol{b})\boldsymbol{a}\boldsymbol{a}^\top)$. However, things are considerably more complicated for $\boldsymbol{D}(\mathbb{A}|\mathbb{B})$ and $\boldsymbol{D}(\boldsymbol{a}|\mathbb{B})$.

---

[6] Lie Product Formula: For any matrices $\boldsymbol{E},\boldsymbol{F}$, $\lim_{n\to\infty}(\mathbf{exp}(\boldsymbol{E}/n)\,\mathbf{exp}(\boldsymbol{F}/n))^n = \mathbf{exp}(\boldsymbol{E}+\boldsymbol{F}) = \mathbf{exp}(\boldsymbol{E})\odot\mathbf{exp}(\boldsymbol{F})$.

[7] Golden-Thompson Inequality: For any symmetric matrices $\boldsymbol{Q}$ and $\boldsymbol{R}$, $\mathrm{tr}(\mathbf{exp}(\boldsymbol{Q}+\boldsymbol{R})\le\mathrm{tr}(\mathbf{exp}(\boldsymbol{Q})\,\mathbf{exp}(\boldsymbol{R}))$.

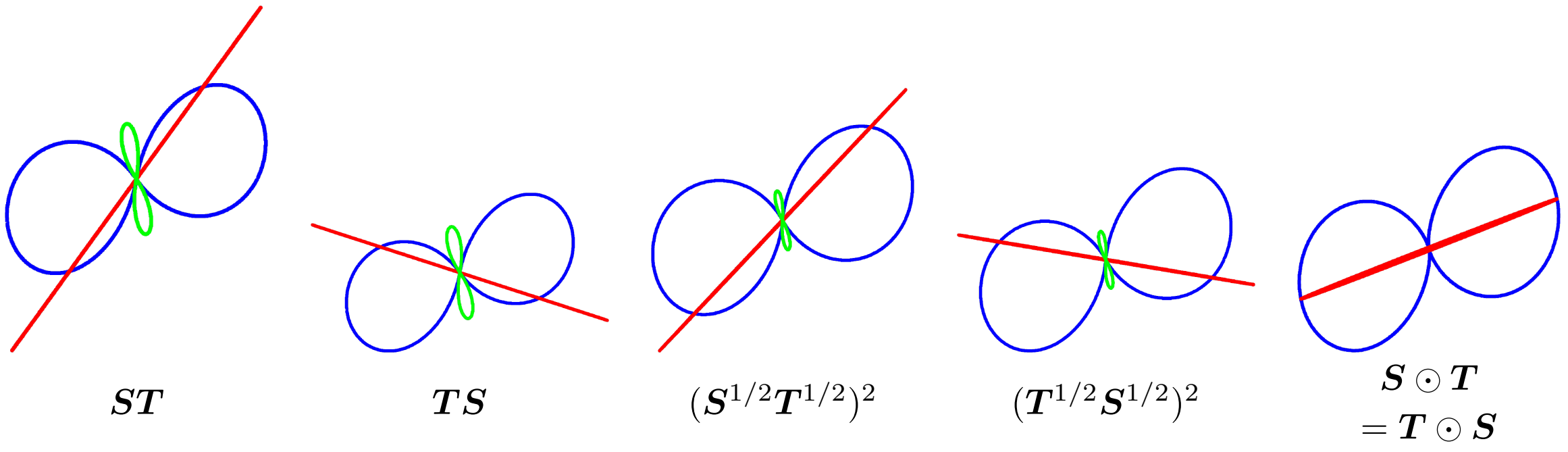

Figure 4: Behavior of the limit formula for $\odot$ operation. We can see that the "ears" indicating negative definiteness are smaller for $(\boldsymbol{S}^{1/2}\boldsymbol{T}^{1/2})^2$ compared to $\boldsymbol{ST}$. As $n$ increases the ears shrink further and $\lim_{n\to\infty}(\boldsymbol{S}^{1/n}\boldsymbol{T}^{1/n})^n = \boldsymbol{S}\odot\boldsymbol{T}$ becomes positive definite. Also, $\boldsymbol{ST}$ and $\boldsymbol{TS}$ are fairly different from one another. The matrices $(\boldsymbol{S}^{1/2}\boldsymbol{T}^{1/2})^2$ and $(\boldsymbol{T}^{1/2}\boldsymbol{S}^{1/2})^2$ are already more similar and the difference between the two multiplication orders decreases with $n$ until in the limit $\boldsymbol{S} \odot \boldsymbol{T} = \boldsymbol{T} \odot \boldsymbol{S}$.

One of the cases for which $\boldsymbol{D}(\mathbb{A}|\mathbb{B})$ can be marginalized is the case of a separable joint density matrix: $\boldsymbol{D}(\mathbb{A}, \mathbb{B})$ is *separable* when it can be written as a weighted sum of Kronecker products, i.e. $\boldsymbol{D}(\mathbb{A}, \mathbb{B}) = \sum_k c_k(\boldsymbol{a}_k \otimes \boldsymbol{b}_k)(\boldsymbol{a}_k \otimes \boldsymbol{b}_k)^\top$. In general it is a difficult problem to determine whether a given joint density matrix is separable. In fact, one of the goals for which [CA99] introduced a conditional density matrix was to give a necessary condition for the separability of a joint density matrix.

For non-separable joints another subtle phenomenon occurs. In the conventional case the mapping from the joint table $P(A, B)$ to the full conditional table $P(A|B)$ is not invertible, since there are infinitely many different joints that give rise to the same conditional. However, we conjecture, based on some numerical experiments, that for non-separable generalized joints $\boldsymbol{D}(\mathbb{A}, \mathbb{B})$ this inverse is in fact unique. There appears to be an EM-like algorithm that converges to $\boldsymbol{D}(\mathbb{B})$ and then $\boldsymbol{D}(\mathbb{A}, \mathbb{B})$ is computed as $\boldsymbol{D}(\mathbb{A}|\mathbb{B}) \odot \boldsymbol{D}(\mathbb{B})$: the estimate $\boldsymbol{W}_{t+1}$ for $\boldsymbol{D}(\mathbb{B})$ is computed from $\boldsymbol{D}(\mathbb{A}|\mathbb{B})$ and the previous estimate $\boldsymbol{W}_t$ as

$$\boldsymbol{W}_{t+1} = \frac{\operatorname{tr}_{\mathbb{A}}(\boldsymbol{D}(\mathbb{A}|\mathbb{B}) \odot (\boldsymbol{I}_{\mathbb{A}} \otimes \boldsymbol{W}_t)}{\operatorname{tr}(\boldsymbol{D}(\mathbb{A}|\mathbb{B}) \odot (\boldsymbol{I}_{\mathbb{A}} \otimes \boldsymbol{W}_t))}.$$

It is easy to see that $\boldsymbol{D}(\mathbb{B})$ is a fixed point of this iteration. We are currently working on convergence and uniqueness proofs. If our conjecture is correct, then we will be able to complete the rules for marginalizing the conditional matrices.

## 7 Theorems of Total Probability

The Theorem of Total Probability is an important calculation in conventional probability theory. It expresses probability of some event $a$ as an expected conditional probability of the elementary events $b_i$ of the other variable:

$$P(a) = \sum_i P(a|b_i)P(b_i).$$

In a more general version, the sum is over any partition $b_i$ of the probability space $B$ into disjoint events. For us the summation becomes a trace and disjoint events become orthogonal vectors.

The first two formulas that are similar to the Theorem of Total Probability follow from the properties of the trace operation:

1. For any orthogonal basis $\boldsymbol{a}_i$ of space $\mathbb{A}$, $\operatorname{tr}(\boldsymbol{D}(\mathbb{A})) = \operatorname{tr}(\boldsymbol{D}(\mathbb{A})\boldsymbol{I}) = \operatorname{tr}(\boldsymbol{D}(\mathbb{A})\sum_i \boldsymbol{a}_i\boldsymbol{a}_i^\top) = \sum_i \boldsymbol{D}(\boldsymbol{a}_i)$.
2. For any orthogonal basis $\boldsymbol{b}_i$ of $\mathbb{B}$, $\boldsymbol{D}(\boldsymbol{a}) = \sum_i \boldsymbol{D}(\boldsymbol{a}|\boldsymbol{b}_i)\boldsymbol{D}(\boldsymbol{b}_i)$.

The second formula can be shown as follows: $\boldsymbol{D}(\boldsymbol{a}) = \operatorname{tr}(\boldsymbol{D}(\boldsymbol{a}, \mathbb{B})) = \sum_i \boldsymbol{b}_i^\top\boldsymbol{D}(\boldsymbol{a}, \mathbb{B})\boldsymbol{b}_i = \sum_i \boldsymbol{D}(\boldsymbol{a}, \boldsymbol{b}_i) = \sum_i \boldsymbol{D}(\boldsymbol{a}|\boldsymbol{b}_i)\boldsymbol{D}(\boldsymbol{b}_i)$.

There are two fancier analogs of the Theorem of Total Probability based on the multiplication operation $\odot$:

1. $\boldsymbol{D}(\boldsymbol{a}) = \operatorname{tr}(\boldsymbol{D}(\boldsymbol{a}|\mathbb{B}) \odot \boldsymbol{D}(\mathbb{B}))$
2. $\boldsymbol{D}(\mathbb{A}) = \operatorname{tr}_{\mathbb{B}}(\boldsymbol{D}(\mathbb{A}|\mathbb{B}) \odot (\boldsymbol{I}_{\mathbb{A}} \otimes \boldsymbol{D}(\mathbb{B})))$.

It is easy to see that these rules are compatible with the formulas for conditionals given in the previous section. As always, the conventional Theorem of Total Probability is obtained when the density and conditional matrices are diagonal.

## 8 Bayes Rules

In the conventional setup we assume that a model $M_i$ is chosen with prior probability $P(M_i)$. The model then generates the data $y$ with probability $P(y|M_i)$, i.e.

$$\begin{aligned} P(y) &= \sum_i P(M_i)P(y|M_i) \\ &= \mathrm{tr}(\mathrm{diag}\,((P(M_i))\,\mathrm{diag}\,((P(y|M_i)))\,. \end{aligned}$$

The reason why we expressed $P(y)$ as a trace of two diagonal matrices will become apparent in a moment.

The generalized setup is completely analogous. There is an underlying joint space $(\mathbb{M}, \mathbb{Y})$ between the model space $\mathbb{M}$ and the data space $\mathbb{Y}$. The prior is specified by a density matrix $\boldsymbol{D}(\mathbb{M})$. The data is a unit direction $\boldsymbol{y}$ in $\mathbb{Y}$ space that is generated by the density $\boldsymbol{D}(\mathbb{Y})$. The probability $\boldsymbol{D}(\boldsymbol{y})$ can be expressed i.t.o. the prior $\boldsymbol{D}(\mathbb{M})$ and data likelihood $\boldsymbol{D}(\boldsymbol{y}|\mathbb{M})$ using the $\odot$ operation:

$$\boldsymbol{D}(\boldsymbol{y}) = \mathrm{tr}(\boldsymbol{D}(\mathbb{M}) \odot \boldsymbol{D}(\boldsymbol{y}|\mathbb{M})).$$

Note that in the conventional case we first chose a model based on the prior and then generated data based on the chosen model. In the generalized case we do not know how to decouple the action on the prior from the choice of the data when conditioned on the prior.

Let us first recall the conventional Bayes rule and rewrite it in matrix notation:

$$P(M_i|y) = \frac{P(M_i)P(y|M_i)}{\sum_j P(M_j)P(y|M_j)} \tag{2}$$

$$\mathrm{diag}\,(P(M_i|y)) = \frac{\mathrm{diag}\,(P(M_i))\,\mathrm{diag}\,(P(y|M_i))}{\mathrm{tr}\,(\mathrm{diag}\,(P(M_i))\,\mathrm{diag}\,(P(y|M_i)))}.$$

We now present and discuss the analogous Bayes rule for the generalized setting. At the end of this section we present a list of all Bayes rules.

In the generalized Bayes rule we cannot simply multiply the prior density matrix with the data likelihood matrix. This is because a product of two symmetric positive definite matrices can be neither symmetric nor positive definite (See Figures 3). Instead, we replace the matrix multiplication with $\odot$ operation:

$$\boldsymbol{D}(\mathbb{M}|\boldsymbol{y}) = \frac{\boldsymbol{D}(\mathbb{M}) \odot \boldsymbol{D}(\boldsymbol{y}|\mathbb{M})}{\mathrm{tr}(\boldsymbol{D}(\mathbb{M}) \odot \boldsymbol{D}(\boldsymbol{y}|\mathbb{M}))}. \tag{3}$$

Normalizing by the trace ensures that the trace of the posterior density matrix is one. In both the conventional as well as the new Bayes rule above, the normalization constant is the likelihood of the data. When the matrices $\boldsymbol{D}(\mathbb{M})$ and $\boldsymbol{D}(\boldsymbol{y}|\mathbb{M})$ have the same eigensystem, then $\odot$ becomes the matrix multiplication. Both of the above Bayes rules can be derived from the minimum relative entropy principle. For the conventional Bayes rule the standard relative entropy between probability vectors is used, whereas the generalized Bayes rule and the crucial $\odot$ operation is motivated by the quantum relative entropy between density matrices due to Umegaki (see e.g. [NC00]). For details of the latter derivation see our previous paper [War05].

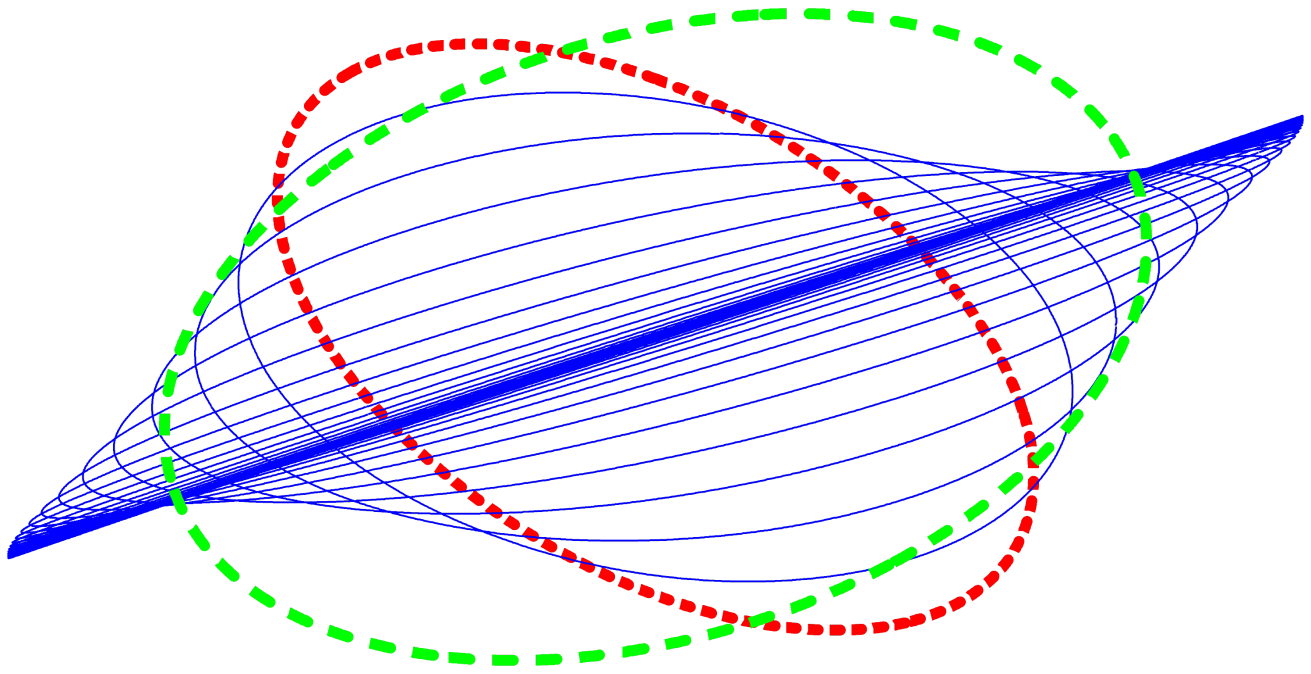

Figure 5: An iterative update of the posterior density matrix using the same likelihood. The likelihood matrix $\boldsymbol{D}(\boldsymbol{y}|\mathbb{M})$ is the big dashed ellipse. The initial prior $\boldsymbol{D}(\mathbb{M})$ is dotted. We can see that the update converges to the largest axis of the likelihood matrix.

As visualized in Figure 5, the posterior moves towards the eigenvector belonging to the largest eigenvalue of the data likelihood matrix $\boldsymbol{D}(\boldsymbol{y}|\mathbb{M})$. Thus the new rule can be interpreted as a soft calculation of the eigenvector with maximum eigenvalue. In the conventional case, the eigensystem stays fixed and the posterior moves towards the maximum component of the data likelihood vector $(P(y|M_i))$. The latter may be seen as a soft max-likelihood calculation.

Other Bayes-like rules for our calculus are listed below. Note that the normalization factors, whether constants or matrices, are always instances of the above Theorems of Total Probability.

1. $\boldsymbol{D}(\mathbb{B}|\mathbb{A}) = (\boldsymbol{I}_{\mathbb{A}} \otimes \boldsymbol{D}(\mathbb{B})) \odot \boldsymbol{D}(\mathbb{A}|\mathbb{B}) \odot$
   $\odot \left(\mathrm{tr}_{\mathbb{B}}\left((\boldsymbol{I}_{\mathbb{A}} \otimes \boldsymbol{D}(\mathbb{B})) \odot \boldsymbol{D}(\mathbb{A}|\mathbb{B})\right) \otimes \boldsymbol{I}_{\mathbb{B}}\right)^{-1}$.

2. $\boldsymbol{D}(\boldsymbol{b}|\mathbb{A}) = \boldsymbol{D}(\boldsymbol{b})\boldsymbol{D}(\mathbb{A}|\boldsymbol{b}) \odot$
   $\odot \left(\mathrm{tr}_{\mathbb{B}}\left(\boldsymbol{D}(\mathbb{A}|\mathbb{B}) \odot (\boldsymbol{I}_{\mathbb{A}} \otimes \boldsymbol{D}(\mathbb{B}))\right)\right)^{-1}$.

3. $\boldsymbol{D}(\mathbb{B}|\boldsymbol{a}) = \dfrac{\boldsymbol{D}(\mathbb{B}) \odot \boldsymbol{D}(\boldsymbol{a}|\mathbb{B})}{\mathrm{tr}(\boldsymbol{D}(\mathbb{B}) \odot \boldsymbol{D}(\boldsymbol{a}|\mathbb{B}))}$.
   This is the Bayes rule derived in [War05] that was discussed above.

4. $\boldsymbol{D}(\boldsymbol{b}|\boldsymbol{a}) = \dfrac{\boldsymbol{D}(\boldsymbol{b})\boldsymbol{D}(\boldsymbol{a}|\boldsymbol{b})}{\sum_i \boldsymbol{D}(\boldsymbol{b}_i)\boldsymbol{D}(\boldsymbol{a}|\boldsymbol{b}_i)}$.
   The summation in the normalization factor proceeds over any orthonormal basis $\boldsymbol{b}_i$.

## 9 Bounds

Recall the following conventional bound for the negative log-likelihood of the data i.t.o. the negative log-likelihood of the MAP estimator:

$$\begin{aligned} -\log P(y) &= -\log \sum_i P(y|M_i)P(M_i) \\ &\leq \min_i(-\log P(y|M_i) - \log P(M_i)). \end{aligned} \tag{4}$$

By properties 3 and 4 of the $\odot$ operation,

$$-\log \boldsymbol{m}^\top \boldsymbol{S}\boldsymbol{m} \leq -\boldsymbol{m}^\top (\mathbf{log}\, \boldsymbol{S})\, \boldsymbol{m},$$

for any unit vector $\boldsymbol{m}$ and symmetric positive definite matrix $\boldsymbol{S}$. Using this inequality we can prove an analogous bound for the generalized probabilities:

$$\begin{aligned} -\log \boldsymbol{D}(\boldsymbol{y}) &= -\log \operatorname{tr}(\boldsymbol{D}(\boldsymbol{y}|\mathbb{M}) \odot \boldsymbol{D}(\mathbb{M})) \\ &\leq \min_{\boldsymbol{m}}(-\log \boldsymbol{m}^\top (\boldsymbol{D}(\boldsymbol{y}|\mathbb{M}) \odot \boldsymbol{D}(\mathbb{M}))\, \boldsymbol{m}) \\ &\leq \min_{\boldsymbol{m}}(-\boldsymbol{m}^\top \mathbf{log}(\boldsymbol{D}(\boldsymbol{y}|\mathbb{M}) \odot \boldsymbol{D}(\mathbb{M}))\, \boldsymbol{m}) \\ &= \min_{\boldsymbol{m}}(-\boldsymbol{m}^\top \mathbf{log}\, \boldsymbol{D}(\boldsymbol{y}|\mathbb{M})\, \boldsymbol{m} - \boldsymbol{m}^T \mathbf{log}\, \boldsymbol{D}(\mathbb{M})\, \boldsymbol{m}). \end{aligned}$$

Intuitively, there are two domains: the probability domain and the log probability domain. The conventional bound (4) can also be written in the probability domain:

$$P(y) \geq \max_i P(M_i)P(y|M_i).$$

However for the generalized probability case, there does not seem to be a simple similar inequality in the probability domain. Throughout the paper we always notice that the matrix operations need to be done in the log domain.

In the conventional case, $P(y)$ is also upper bounded by $\max_i P(y|M_i)$. For the generalized case, the analogous formula is the following, where $\mu_i$ and $\boldsymbol{m}_i$ are the eigenvalues/vectors of $\boldsymbol{D}(\mathbb{M})$:

$$\begin{aligned} \boldsymbol{D}(\boldsymbol{y}) &= \operatorname{tr}(\boldsymbol{D}(\boldsymbol{y}|\mathbb{M}) \odot \boldsymbol{D}(\mathbb{M})) \\ &\leq \operatorname{tr}(\boldsymbol{D}(\boldsymbol{y}|\mathbb{M})\boldsymbol{D}(\mathbb{M})) \\ &= \sum_i \mu_i \boldsymbol{m}_i^\top \boldsymbol{D}(\boldsymbol{y}|\mathbb{M})\, \boldsymbol{m}_i \\ &\leq \boldsymbol{m}_j^\top \boldsymbol{D}(\boldsymbol{y}|\mathbb{M})\, \boldsymbol{m}_j, \end{aligned}$$

where $j \in \operatorname{argmax}_i \boldsymbol{m}_i^\top \boldsymbol{D}(\boldsymbol{y}|\mathbb{M})\, \boldsymbol{m}_i$.

## 10 Conclusions

Density matrices are central to quantum physics. We utilize many mathematical techniques from that field to develop a Bayesian probability calculus for density matrices. Intuitively, the new calculus will be useful when the data likelihood $\boldsymbol{D}(\boldsymbol{y}|\mathbb{M})$ has non-zero off-diagonal elements, i.e. information about which components are correlated or anti-correlated. The main new $\boldsymbol{S} \odot \boldsymbol{T}$ operation first takes logs of the matrices $\boldsymbol{S}$ and $\boldsymbol{T}$, adds the logs and finally exponentiates. Any straightforward implementation of the $\odot$ operation requires the eigendecompositions of the matrices, which are expensive to obtain. Throughout our work we notice that the log domain seems to be more important in the matrix case.

We previously derived the $\odot$ operation by employing the minimum relative entropy principle. We now express the main Bayes rules as solutions to certain differential equations and contrast these equations with Schrödinger's equation, the main differential equation of quantum physics. In the conventional case, the differential equations are $(1 \leq i \leq n)$:

$$\frac{\partial \log P(M_i|t)}{\partial t} = \log P(y|M_i) - \sum_j P(M_j|t) \log P(y|M_j).$$

The solution is

$$P(M_i|t) = \frac{P(M_i|0)P(y|M_i)^t}{\sum_j P(M_j|0)P(y|M_j)^t}.$$

If we take the value $P(M_i|0)$ as the prior $P(M_i)$ then the expression for $P(M_i|1)$ becomes the conventional Bayes rule (2). There is a similar differential equation for the generalized Bayes rule (Here we assume that the prior $\boldsymbol{D}(\mathbb{M})$ and data likelihood matrix $\boldsymbol{D}(\boldsymbol{y}|\mathbb{M})$ are strictly positive definite):

$$\frac{\partial\, \mathbf{log}\, \boldsymbol{D}(\mathbb{M}|t)}{\partial t} = \mathbf{log}\, \boldsymbol{D}(\boldsymbol{y}|\mathbb{M}) - \operatorname{tr}(\boldsymbol{D}(\mathbb{M}|t)\, \mathbf{log}\, \boldsymbol{D}(\boldsymbol{y}|\mathbb{M})).$$

The solution has the form

$$\begin{aligned} \boldsymbol{D}(\mathbb{M}|t) &= \frac{\mathbf{exp}\,(\mathbf{log}\, \boldsymbol{D}(\mathbb{M}|0) + t\, \mathbf{log}\, \boldsymbol{D}(\boldsymbol{y}|\mathbb{M}))}{\operatorname{tr}(\mathbf{exp}\,(\mathbf{log}\, \boldsymbol{D}(\mathbb{M}|0) + t\, \mathbf{log}\, \boldsymbol{D}(\boldsymbol{y}|\mathbb{M})))} \\ &\overset{(1)}{=} \frac{\boldsymbol{D}(\mathbb{M}|0) \odot \boldsymbol{D}(\boldsymbol{y}|\mathbb{M})^t}{\operatorname{tr}(\boldsymbol{D}(\mathbb{M}|0) \odot \boldsymbol{D}(\boldsymbol{y}|\mathbb{M})^t)}. \end{aligned}$$

If we set $\boldsymbol{D}(\mathbb{M}|0)$ to the prior $\boldsymbol{D}(\mathbb{M})$, then the expression for $\boldsymbol{D}(\mathbb{M}|1)$ becomes the generalized Bayes rule (3). Notice again that the differential equations emphasize the log domain and that the $\odot$ operation appears in the solution.

For contrast, the main differential equation for density matrices in quantum physics is the following version of the Schrödinger equation [Fey72]:

$$\frac{\partial \boldsymbol{D}(\mathbb{M}|t)}{\partial t} = i\,(\boldsymbol{H}\, \boldsymbol{D}(\mathbb{M}|t) - \boldsymbol{D}(\mathbb{M}|t)\, \boldsymbol{H}),$$

where $\boldsymbol{H}$ is Hermitian. The solution has the form

$$\boldsymbol{D}(\mathbb{M}|t) = \mathbf{exp}(-i\, t\, \boldsymbol{H})\, \boldsymbol{D}(\mathbb{M}|0)\, \mathbf{exp}(i\, t\, \boldsymbol{H}),$$

where $\boldsymbol{D}(\mathbb{M}|0)$ is the initial density matrix. Since $it\boldsymbol{H}$ is skew Hermitian, both exponentials are unitary. Thus the above update represents a unitary transformation of the initial density matrix $\boldsymbol{D}(\mathbb{M}|0)$. Such transformations leave the eigenvalues unchanged and only affect the eigensystem. In contrast our generalized Bayes rule updates both the eigenvalues and eigenvectors, and the conventional Bayes rule can be seen as only updating the eigenvalues while keeping the eigenvectors fixed.

Interestingly enough the $\odot$ operation has also been employed in computer graphics for combining affine transformation [Ale02]. Also the simulation of quantum computations based on the Lie Trotter Formula ([NC00], Chapter 4.7) can be interpreted as applying the $\odot$ operation to unitary matrices and not to density matrices.

At this point we have no convincing application for the new probability calculus. However, a similar methodology was used to derive and prove bounds for parameter updates of density matrices that led to a version of Boosting [TRW05] where the distribution over the examples is replaced by a density matrix, an online variance minimization algorithm where the parameter space is the unit ball [WK06a], and an on-line algorithm for Principal Component Analysis [WK06b].

The new calculus seems to be rich enough to bring out some of the interesting phenomena of quantum physics, such as superposition and entanglement. Maybe the new calculus can be used to maintain "uncertainty" in quantum computation.

Finally, we will reason in a simple case that generalized probability space is more "connected" and a clever algorithm might be able to exploit this. Assume zero is encoded as the distribution $(1,0)$ and one as the distribution $(0,1)$. Moving from the zero distribution to the one distributions can be done by lowering the probability of the first component and increasing the probability of the second. As density matrices, zero and one would be $\left(\begin{smallmatrix}1&0\\0&0\end{smallmatrix}\right)$ and $\left(\begin{smallmatrix}0&0\\0&1\end{smallmatrix}\right)$, respectively. Note that the eigensystem for both matrices is the identity matrix and there is now a second way to go from zero to one that keeps the eigenvalues/probabilities fixed but swaps the eigenvectors:

$$\begin{pmatrix}0&1\\1&0\end{pmatrix}\begin{pmatrix}1&0\\0&0\end{pmatrix}\begin{pmatrix}0&1\\1&0\end{pmatrix}=\begin{pmatrix}0&0\\0&1\end{pmatrix}.$$

**Acknowledgments:** Thanks to Joel Yellin, Thorsten Ehrhardt and anonymous referees for helpful insights.